\documentclass{aa}
\usepackage{graphicx}
\begin{document}
   \title{ Star formation rates of distant luminous infrared
            galaxies derived from H$\alpha$ and IR 
            luminosities\footnote{Based on observations
            with ISO  , HST , CFHT and ESO/VLT  Telescopes} }


   \author{H. Flores\inst{1}
          \and
          F. Hammer\inst{1}
          \and
          D. Elbaz\inst{2}
          \and
          C.J. Cesarsky\inst{3}
          \and
          Y.C. Liang\inst{1}
          \and
          D. Fadda\inst{4}
          \and
          N. Gruel\inst{1}
          }

   \offprints{hector.flores@obspm.fr}
\authorrunning{H. Flores et al.}
\titlerunning{Optical follow-up of IS0 galaxies }

   \institute{GEPI, Observatoire de Paris Meudon, 92190 Meudon, France
              \email{hector.flores@obspm.fr}
         \and
       CEA Saclay - Service d'Astrophysique, Orme des Merisiers, F91191 Gif-sur-Yvette Cedex, France
        \and
       ESO, Karl-Schwarzschild Stra\ss e 2, D85748 Garching bei M\"unchen, Germany
        \and
Instituto de Astrof\'\i sica de Canarias (IAC), Via Lactea S/N, E-38205 La Laguna, Tenerife, Spain   
              }

   \date{---}

   \abstract{We present a study of the star formation rate (SFR) for a
     sample of 16 distant galaxies detected by ISOCAM at 15$\mu$m in
     the CFRS0300+00 and CFRS1400+52 fields. Their high quality and  
     intermediate resolution VLT/FORS spectra have allowed
     a proper correction of the Balmer emission lines from the
     underlying absorption.  Extinction estimates using the
     H$\beta$/H$\gamma$ and the H$\alpha$/H$\beta$ Balmer decrement
     are in excellent agreement, providing a robust measurement of the
     instantaneous SFR based on the extinction-corrected $H\alpha$
     luminosity. Star formation has also been estimated exploiting the
     correlations between IR luminosity and those at MIR and radio
     wavelengths. Our study shows that the relationship between the
     two SFR estimates follow two distinct regimes: (1) for galaxies
     with SFR$_{IR}$ below $\sim 100M_{\odot}/yr$, the SFR deduced from
     H$\alpha$ measurements is a good approximation of the global SFR
     and (2) for galaxies near of ULIRGs regime, corrected H$\alpha$
     SFR understimated the SFR by a factor of 1.5 to 2.  Our 
     analyses suggest that heavily extincted regions completely hidden in
     optical bands (such as those found in Arp 220) contribute to less
     than 20\% of the global budget of star formation history up to
     z=1.   }

   \maketitle
%

\section{Introduction}
In the last few years, great steps toward a better understanding of
the distant Universe have been done using a panchromatic approach.
The mid IR windows opened by the camera ISOCAM (Cesarsky et al, 1996)
have lead to the identification of the galaxies responsible for  the
bulk of the infrared background, a population of dusty galaxies which evolves very rapidly in number (Elbaz et al., 1999, 2002). 
Luminous Infrared Galaxies (LIRGs, L$_{IR}\ge 10^{11}L_{\odot}$)
detected by ISO are mainly dusty starbursts (Fadda et al., 2002) with
large star formation ($> 50-100 M_{\odot}$/yr) triggered by
interactions, $\sim$30-40\% of detected galaxies show optical signs of
interaction (Flores et al., 1999).
Rigopoulou et al. (2000, hereafter R00) using the VLT studied the
near IR spectra of a few LIRGs and claimed that the SFR measured from
IR luminosity (hereafter $SFR_{IR}$) is at least 3 times higher than
that based on $H\alpha$ luminosity (hereafter SFR$_{H\alpha}$).
However, this claim is affected by large uncertainties on the
SFR$_{H\alpha}$, in absence of a proper extinction correction. Indeed
extinction for such heavily dust-shrouded objects is very important,
and only accurate estimates of the extinction can confirm or infirm
the R00 claim.  Recently, Hammer et al. (2001) have
analyzed intermediate resolution VLT/FORS spectra (R=3.5\AA$\;$ at
rest) of 3 compact LIRGS at z$\sim$0.6.  They derived accurate fluxes
for Balmer emission lines (H$\beta$, H$\gamma$) after a proper removal
of the underlying stellar absorption. Extinction estimates using
observed H$\beta$/H$\gamma$ ratio were used to correct $H\beta$
luminosities and derive SFR$_{H\alpha}$ which are consistent with
$SFR_{IR}$.
These objects present strong
metallic and Balmer absorption lines combined with intense emission
lines, which indicate a particularly complex star formation history 
(Hammer et al, 2001). It is unclear yet if these properties are shared
by the whole galaxy population at intermediate redshifts. Indeed, at z $\ge$
0.4, an important fraction of the galaxy population was identified to
Balmer-strong galaxies with emission, in the field (Hammer
et al, 1997) and in galaxy clusters (Poggianti et al, 1999).    \\
This letter presents a preliminary study of 16 LIRGs out of a sample
of 90 observed at VLT and CFHT.  The data are described in Section 2.
Section 3 summarizes the different techniques used for the data
analysis.  The determination of the SFR is discussed in Sections 4 and
5. Throughout this paper we adopt H$_0$=50 km/s/Mpc and q$_0$=0.5 ($\Omega_M=1,\Omega_{Lambda}=0$).

\begin{figure*}   
   \centering \includegraphics[width=12cm]{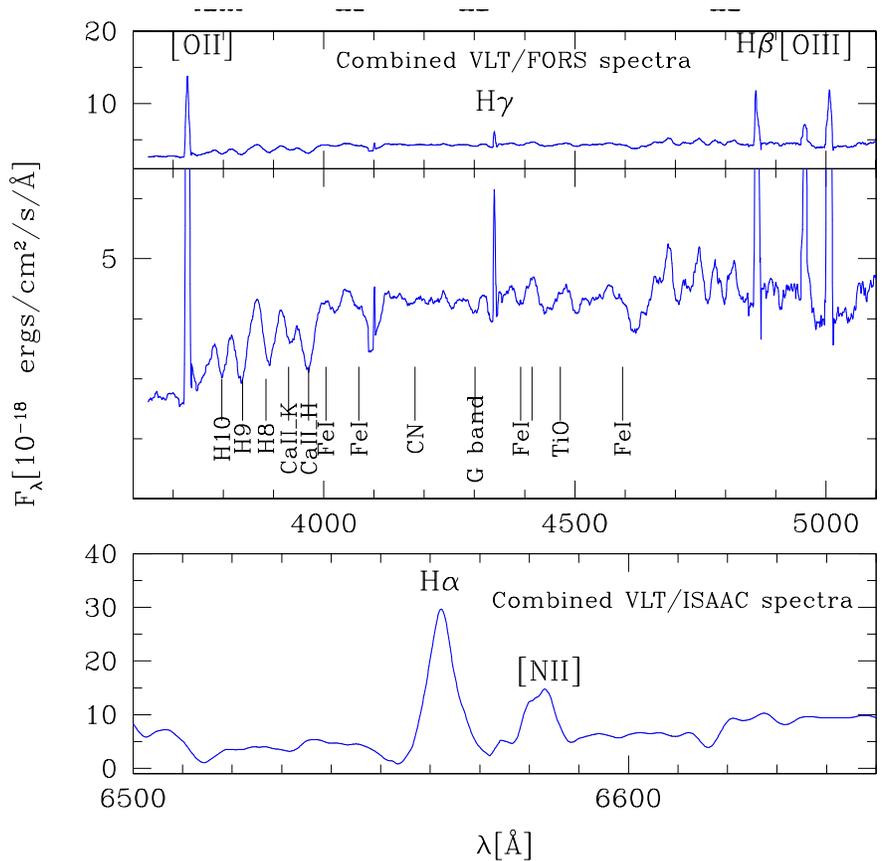} 
\caption{ 
  Combined VLT FORS+ISAAC spectra, after masking sky emission lines,
  and smoothing the continuum of both galaxies (see Hammer et al.
  2001;Gruel et al. in prep).  Zooms in the center shows details of
  our spectra (S/N$>$5 per resolution element in the continuum),
  including the underlying Balmer absorption.  Lower panel shows
  combined near IR spectra.
}
   \label{Fig1}
\end{figure*}
\section{Observations and sample selection}
Deep MIR data  were obtained in a region of
13x13 sq. arcmin centered on the CFRS0300+00 field, using ISOCAM
broadband filters LW3 (12-18\,$\mu$m, centered at 14.3\,$\mu$m) and LW2
(5-8\,$\mu$m, centered at 6.75\,$\mu$m) for a total integration time per
sky position of 23 and 15 minutes, respectively. The reduction of IR
data was done using the standard method described by Fadda et al
(2001), resulting in a catalog of detected sources, which will be
published in a subsequent paper. 

Among the 40 LIRGs galaxies possessing a redshift (from Canada France
Redshift Survey, Hammer et al, 1995) we have randomly selected 14
galaxies which have been observed together using the VLT/FORS2/MXU
(runs 66.A-0599(A), 68.A-0298(A)) and 2 additional objects were
observed with the CFHT/MOS (run 98IF65A). The 8 distant LIRGs (z$>$
0.6) were pre-selected to avoid contamination of their near IR
redshifted $H\alpha$ line by strong OH sky emissions. The sample is
then well representative of LIRGs in the CFRS0300+00 field.  Optical
VLT observations were done in service mode during periods 66 and 68
using two differents FORS2 set-ups (R600 and I600 grisms, 3 hour each,
R$\sim$1300) with a slit width of 1.5 arcsec. CFHT data were obtained
in two different runs in 1998 and 2000, using standard MOS setup (R300
grism and 1.5 arcsec slit, 4 hrs, R=660).  Near IR spectra of four
galaxies were collected in runs 66.A-0599(A) and 69.B-0301(A) with the
infrared spectrometer VLT/ISAAC.  For the near IR observations we used
the medium resolution grating R$_{s} \sim $ 3000 and a 2'' slit, for a
total of 1 hr of integration time nodding along the slit (ABBA
configuration).  Targets were first acquired using a reference star at
a distance of 1', with a 1--2 min exposure in the Js-band.  For four
objects, we have used VLT/ISAAC archive data (Program 63.O-0270(A)).
These galaxies were observed using the same instrumental setup.  Data
reduction and extraction of optical and near-IR spectra were performed
using a set of IRAF procedures developed by our team, which allowed us
to reconstruct simultaneously the spectra and the sky counts of the
objects.

\section{Data analysis procedure}
Figure 1 shows the median optical (VLT/FORS) and the near IR VLT
spectra (VLT/ISAAC) of the 8 distant LIRGs. The spectra were
flux-calibrated using standard stars for both optical and near IR
spectra. Broadband filter images (V to I bands for FORS spectra, J
band for ISAAC spectra) were used to compute the spectroscopic
aperture corrections. In order to check the validity of the relative
aperture corrections, we have produced best fit models of both spectra
and photometry using population synthesis models (Bruzual and Charlot,
1993) appropriately corrected for extinction.  This procedure has
provided an independant test of the spectral calibration and of
possible errors in aperture corrections. It results that the
uncertainty related to our derived calibration is well below 15\%.
Underlying Balmer absorption line were estimated after synthetizing
the galaxy continuum from 3730\AA~ to 5000\AA, including major
metallic lines and the high order absorption Balmer lines unaffected
by emissions.  Synthetic spectra were produced from combinations of 4
stars from B to K type, selected from Jacoby et al (1984),
appropriately chose to best fit the continuum and absorption line
spectra (see details in Hammer et al, 2001; Gruel PhD thesis, 2002).
   \begin{figure}
   \centering
   \includegraphics[width=6cm]{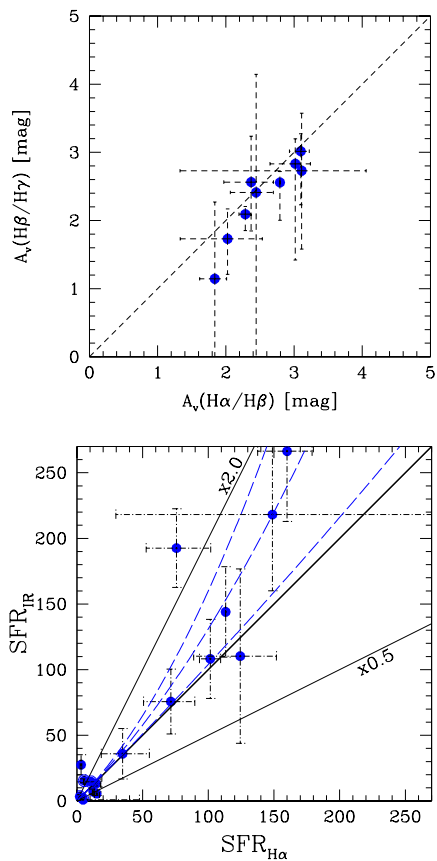}
   \caption{ (a){\it Upper panel: } This plot shows the strong 
     correlation that exists between the extinction estimated using the
     two differents Balmer ratios (H$\beta$/H$\gamma$ and
     H$\alpha$/H$\beta$). (b) {\it Bottom panel: } Comparison of the
     two SFRs, using MIR luminosities and Balmer emission line
     luminosities. Full line represents SFR$_{IR}$=(2-1-0.5) $\times$
     SFR$H\alpha$. The dashed-line represents a polynomial (best) fit to
     the points (SFR$_{IR}$= SFR$_{H\alpha}
     \times$(1+a$\times$SFR$_{H\alpha}$), where a=3.11$^{6.0}_{0.4}\times
     10^{-3}$). SFR$_{H\alpha}$ underestimates the global SFR for
     values larger than $\sim$100M$_{\odot}$/yr.  }
   \label{Fig2}
    \end{figure}

\begin{table*}
      \caption[]{Extinctions and SFR of distant LIRGs and nearby galaxies d
etected by ISO( see enclosed jpg).}
         \label{raies2}
\end{table*}

 \section{Extinction and SFR from $H\alpha$, $H\beta$ and $H\gamma$ Balmer emission lines} 
Median resolution VLT spectra (FORS+ISAAC) allowed us to measure three
Balmer emission lines (H$\gamma$, H$\beta$ and H$\alpha$) for distant
LIRGs. Extinctions were derived from both the H$\beta$/H$\gamma$ and
the H$\alpha$/H$\beta$ Balmer ratios using standard extinction law
(Fitzpatrick, 1999)
Table 1 lists the corresponding $A_{V}$ and the associated error bars.
Relative errors due to S/N and data analysis methodology have then
been quadratically added, and hence the final error bars correspond to
more than a 1$\sigma$ error.  The extinction factor of LIRGs averages
to $A_{V}$ $\sim$ 2.8 at z$\sim$0.7, a value much larger than
$A_{v}$=1.12 obtained by Kennicutt (1992) for normal spiral galaxies
or $A_{V}$=0.57 derived by Gallagher et al. (1989) for irregular
galaxies.  Figure 2 shows the excellent correlation between extinction
estimates from the H$\beta$/H$\gamma$ and the H$\alpha$/H$\beta$
ratios.  SFR were calculated using extinction corrected H$\alpha$ line
(H$\beta$ in the case of galaxy 03.0495) integrated flux following the
recipes of Kennicutt (1998).  The correlation displayed in Figure 2a
is valid over a wide range in redshift and luminosity. We give in this
paper the cleanest evaluation of SFR of starbursts and LIRGS done up
to now on the basis of Balmer lines. Having reached the limits of the
method, we evaluate the validity of its application for ascertaining
actual SFRs in dusty galaxies.
\section{ SFR derived from multi-wavelength correlations}
A major fraction of the ionising photons related to star formation
were reprocessed by dust in dust-enshrouded galaxies such as the LIRGs
studied here.  The IR luminosity (8--1000 $\mu$m) may also provide a
good estimate of the SFR. We estimated the SFR using  a set
of SED template from the STARDUST II model (Chanial et al, in prep) and 
the relationship found by Elbaz et al. (2002) between the MIR and the
IR luminosity.  For a few objects detected at radio wavelengths we
have also derived the IR luminosity from the radio-FIR correlation
(see Condon, 1992).  All three methods produce results consistent
within the error bars which are dominated by the uncertainties in the
MIR photometry and the scatter of MIR (or radio)- IR relationship.
SFR has been computed assuming the recipe of Kennicutt (1998), and
allowing a direct comparison with our derivation of the SFR from
Balmer emission lines, since both calculations assumed a Salpeter
(1995) IMF with mass limits from 0.1 to 100$M_{\odot}$.
\section{Discussion and conclusions}
Extinction effects potentially affect a large cosmological sources and
their consequences have been the subject of a long debate in
observational cosmology. 
In this letter, we study a representative
sample of dust enshrouded starbursts, which, at moderate redshifts,
can be extensively studied in spectroscopy, at MIR and at radio
wavelengths. We have compared the instantaneous SFR related to the
optical nebular emission (SFR$_{H\alpha}$) to the SFR estimated from
the MIR emission measured by ISO 
, and, when available, from the radio emission
These two methods give compatible results, and we denote the result by
(SFR$_{IR}$).  H$\alpha$ luminosities have been properly corrected
for:(1) extinction corrections (using Balmer decrement ratio); (2)
absorption line contamination (using moderate resolution spectroscopy)
and (3) aperture correction (using photometric measurements). The
latter correction could potentially lead to major errors. Indeed it
assumes that the fraction of the emission line light which is sampled
by the slit equals that of the continuum light derived from broad band
photometry, which could be wrong.  However, our extinction estimate
from the H$\beta$/H$\gamma$ ratio does not depend on such assumption
and agrees well with our estimate from the H$\alpha$/H$\beta$ ratio.
Moreover, for the 8 distant galaxies, the aperture correction is
generally close to 1, and so we believe that uncertainties related to
aperture correction are marginal.
IR luminosities have been estimated by (1) using the correlation between
MIR and far-IR luminosities and (2) using the correlation between
radio and far-IR luminosities. A good agreement is found using the two
methods.  Despite the large error bars on individual SFRs, we find a
remarquable trend in the relationship between the two SFR estimates:
 (1) For starbursts with SFR$_{IR}$ below
  $90-130M_{\odot}/yr$, SFR estimates based on H$\alpha$ luminosities,
  properly corrected,
  agree well with SFR$_{IR}$, in agreement with analyses of local
  galaxies with low SFRs (Buat et al., 2002, Rosa-Gonzalez et al.,
  2002);
(2) At larger IR luminosities, ${H\alpha}$ emission lines
  underestimate the SFR by factors up to 2.5  at
  SFR$_{IR}$=$250M_{\odot}/yr$.
 (3) few objects show properties strongly discrepant from the
  empirical relationship between SFR$_{H\alpha}$ and SFR$_{IR}$ which
  could be due to their individual star formation history: for
  example, if the star formation was rapidly decreasing (increasing),
  one would expect SFR$_{IR}$ larger (smaller) than SFR$_{H\alpha}$.
  Our result is in excellent agreement with the recent study of the
  IRAS sources detected by the
  SDSS (Hopkins et
  al, 2003, see their Fig 15) in the local universe. However it
  substantially differs from that of Franceschini et al (2003,
  hereafter F03), who have studied similar sources in the 
  HDFS. Indeed in the F03's Fig. 8, IR and $H\alpha$ estimates
  of the SFR are almost similar for LIRGs and ULIRGs, while for most
  of the less luminous IR galaxies, infrared provide larger values of
  the SFRs. Although the above results are somewhat in contradiction
  with our and Hopkins et al studies, one has to understand the major
  reasons of such a discrepancy.
  F03's Table 5 displays extinction corrected
  SFR(Halpha) values for 11 objects. For 7 of them, they have
  estimated dust extinction from the stellar continuum on the basis of
  the rest-frame V-K colors or on the comparison between H$\alpha$ and
  rest-frame 2800AA fluxes; it is generally believed that the
  continuum can only provide gross estimates of the gas extinction,
  because it is strongly affected by the various and complex star
  formation histories.  For the 4 resting objects (HDFS25, 27, 53 and
  55), the discrepancy might be related to the fact that F03
   have not corrected H$\beta$ fluxes for underlying absorption,
  leading to a systematic underestimate of H$\beta$ emission flux.
  Moreover 2 of these objects (HDFS25, 55) present a low S/N for the
  H$\alpha$ line (see R00 Fig. 1).
  
  Using both SFR's, we have estimated an empirical polynomial
  relationship SFR$_{IR}$ = SFR$_{H\alpha}
  \times$ (A$\times$ SFR $_{H\alpha}$ + 1), where A=3.11$^{6.0}_{0.4}$
  $10^{-3}$ (which fits at best the points in Figure 2b, weighted
  according to the error bars).  Applying our empirical relationship
  to the L$_{IR}$ estimated by Flores et al.  (1999) on field
  galaxies, one can derive in principle the H$\alpha$ luminosity
  density corresponding to the IR luminosity density at z $\le$ 1
  (note that this result is insensitive to a change of cosmological
  constants). It results that the $H\alpha$ star formation density
  corresponds to 83\% to 89\% of the IR star formation density, for
  the Flores et al. lower and upper limit, respectively.  This suggest
  that galaxies which actually dominates the cosmic star formation
  density at z$\le$ 1 are not ULIRGs, but correspond to the much more
  abundant population of LIRGs and starbursts.  
   For determining the SFR, MIR
  fluxes which can be measured rapidly by satellites (ISO, SIRTF)
  give, and at least for the intrinsically bright IR galaxies, more
  accurate estimates than cumbersome optical and NIR spectroscopy.
  However, the detailed spectra bring useful information on metal
  abundances and on the evolutionary history of the galaxies, as we
  will show in subsequent papers.

{\it Acknowledgements:} 
We thank Rafael Guzm\'an for enlightening discussions which greatly
contributed to improve our paper. 



\begin{thebibliography}{}
\bibitem[]{} Buat et al..,  2002A\&A, 383, 801.
\bibitem[]{} Cesarsky, C. et al., 1996, A\&A 315, 32.
\bibitem[]{} Elbaz, D., et al.., 1999, A\&A 351, L37
\bibitem[]{} Elbaz, D., et al. ,2002, A\&A 384, 848.
\bibitem[]{} Fadda, D., et al.., 2000, A\&A 361, 827.
\bibitem[]{} Fadda, D., et al.., 2002, A\&A 383, 838.
\bibitem[]{} Fitzpatrick, E., 1999, PASP, 111, 63.
\bibitem[]{} Flores, H., et al., 1999, ApJ, 517, 148.
\bibitem[]{} Franceschini, A., 2003, A\&A 403, 501
\bibitem[]{} Gallagher, J.Hunter, D. A., Bushouse, H., AJ 98, 806.
\bibitem[]{} Hammer, F., et al., 1997, ApJ 481, 49.
\bibitem[]{} Hammer, F., et al., 2001, ApJ 550, 570.
\bibitem[]{} Hopkins, A. M. et al, 2003, astro-ph/0306621
\bibitem[]{} Kennicutt, R.C., 1998, Ann. rev. Astr. Ap. 36, 189
\bibitem[]{} Kennicutt, R.C., 1992 ApJ, 388, 310.
\bibitem[]{} Poggianti, B.M., et al., 1999, ApJ, 518, 576
\bibitem[]{} Rigopoulou, D., et al., 2000, ApJ 537, 85.
\bibitem[]{} Rosa-Gonzalez et al., 2002, MNRAS, 332, 283.
\end{thebibliography}
\end{document}